\documentstyle[psfig,times]{mn}
 
\def\xmm{{\it XMM-Newton}}

\def\et{{et al.\ }}
\def\hst{{\it HST~\/}}

\def\asca{{\it ASCA~\/}}

\def\gsim{\mathrel{\hbox{\rlap{\hbox{\lower4pt\hbox{$\sim$}}}{\raise2pt\hbox{$>$}}}}}
\newcommand{\ls}{\mathrel{\hbox{\rlap{\hbox{\lower4pt\hbox{$\sim$}}}\hbox{$<$}}}}
\newcommand{\gs}{\mathrel{\hbox{\rlap{\hbox{\lower4pt\hbox{$\sim$}}}\hbox{$>$}}}}
     
\def\arcs{{\hbox{$^{\prime\prime}$}}}

\def\H0{{\rm ~km~s^{-1}~Mpc^{-1}}}

\def\et{{et al.~\/}}

\def\deg{^\circ}
\def\feii{{Fe~\textsc{ii}}}
\def\hb{{H$\beta$}}

\def\1{{1H~0707--495}}

\title[\xmm\ discovery of a sharp spectral feature in \1 ]
{\xmm\ discovery of a sharp spectral feature at $\sim$7~keV in the 
Narrow-Line Seyfert~1 galaxy \1}

\author[Th. Boller \et]
{Th. Boller,$^1$ A. C. Fabian,$^2$ R. Sunyaev,$^3$ J. Tr\"umper,$^1$ 
S. Vaughan,$^2$ D. R.~Ballantyne,$^2$
\newauthor W. N. Brandt,$^4$ R. Keil,$^1$ and K. Iwasawa$^2$\\
$^1$Max-Planck-Institut f\"ur extraterrestrische Physik, Postfach 1603, 85748 Garching, Germany\\
$^2$ Institute of  Astronomy, Madingley Road, Cambridge CB3 0HA\\
$^3$ Max-Planck-Institut f\"ur Astrophysik, 85748 Garching, Germany\\
$^4$ Department of Astronomy and Astrophysics, 
The Pennsylvania State University, 525 Davey Lab, University Park, PA 16802, USA
}
\date{draft 12/10/2001}
\pagerange{\pageref{firstpage}--\pageref{lastpage}}
\pubyear{2001}
\begin{document}

\maketitle
\label{firstpage}
\begin{abstract}
We report the first detection of a sharp spectral feature  in a
Narrow-Line Seyfert 1 galaxy.  Using \xmm\ we have observed \1 and
find a drop in flux by a factor  of more than 2  at a rest-frame
energy of $\sim$7~keV  without any detectable narrow Fe K$\alpha$
line emission.  The energy of this feature suggests a connection with
the neutral  iron K photoelectric edge,  but the lack of any obvious
absorption in the spectrum at lower energies  makes the interpretation
challenging.  We explore two alternative explanations for this unusual
spectral feature: (i) partial covering absorption by clouds of neutral
material and (ii) ionised disc reflection with  lines and edges from
different ionisation stages of iron blurred together by relativistic
effects.
We note that both models require an iron overabundance to explain the
depth of the feature.   The X-ray light curve shows strong and rapid
variability, changing by a factor of four during the observation.
The source displays modest spectral variability which is uncorrelated
with flux. 
\end{abstract}

\section{Introduction}
\label{sect:intro}

\1 ($B=15.7$; $z=0.0411$) has displayed some of the most rapid X-ray
variability observed in a Seyfert galaxy (Turner \et 1999; Leighly
1999a).  Its X-ray spectrum is steep and shows a strong ``soft
excess'' (Vaughan \et 1999;  Leighly 1999b) and its optical spectrum
shows narrow \hb\ (FWHM \hb~= 1050 km s$^{-1}$) and strong \feii\
emission (Remillard et al. 1986, Leighly 1999b), characteristic of many ultrasoft
Seyferts. On the basis of FWHM \hb\ \1 is classified as a Narrow-Line
Seyfert 1 (NLS1) galaxy.

Such ultrasoft NLS1s are of particular interest because their extreme
properties are most likely produced by an extreme value of an
underlying physical parameter (e.g. Pounds, Done \& Osborne 1995; Boller, Brandt \&
Fink 1996; Brandt \& Boller 1998; Vaughan \et 2001). However, with a few
exceptions, their timing and spectral properties proved difficult to study in
detail with previous X-ray observatories as a result of e.g.,  regular light
curve interruptions from Earth occultation and limited spectral
bandpass.

\1 was observed as part of a guaranteed time programme to  study the
timing and spectral properties of the X-ray brightest and most
prominent ultrasoft NLS1s using \xmm.  In this Letter we report the
first results from the EPIC data and in particular the detection of an
unusual spectral feature at around 7~keV.

\section{Observation and data reduction}
\label{sect:obs}

\begin{figure*}
\hspace{0.3 cm}
\psfig{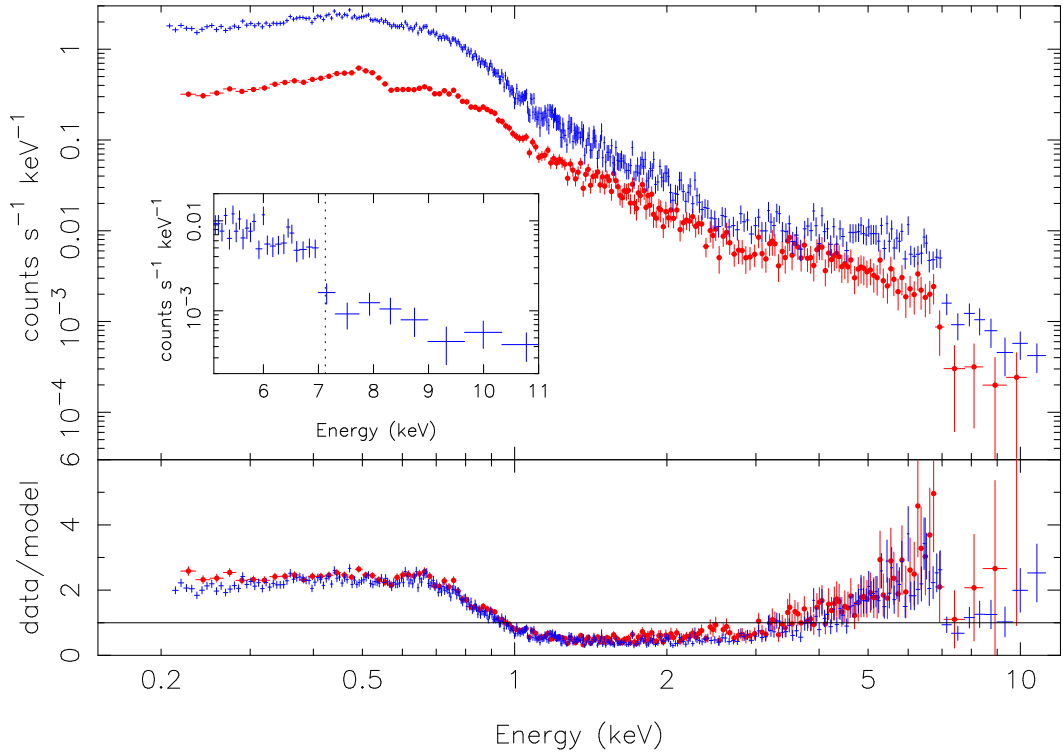}
      \caption{
Raw EPIC pn (blue) and MOS (red) count spectra of \1 shifted to the
rest-frame of the source. (The MOS1 and MOS2 spectra were combined for
display purposes only.) These spectra clearly show a significant drop at
$\sim 7$~keV. The inset panel shows the pn spectrum, again in the
rest-frame of the source, with the expected position of the neutral iron K
edge marked (dotted line). 
The bottom panel shows the data compared to a
power-law model, with the slope fixed at $\Gamma=2.7$. The model was not
fitted to the data but chosen to emphasise the shape of the spectrum at
both soft and hard energies.
\label{fig:raw_spec}}
\end{figure*}

\xmm\ observed \1 on 2000 October 21 during rev. 0159 for a duration
of 46 ks, during which all instruments were operating nominally.
The EPIC pn camera was operated in full-frame mode, and the two MOS
cameras were in large-window mode;  all three cameras used the medium
filters. Extraction of science products from the Observation Data Files
(ODFs) followed standard procedures using the \xmm\ Science Analysis
System version 5.1.0 (SAS\footnote{See {\tt
http://xmm.vilspa.esa.es/user/sas\_top.html}}). 

The raw MOS and pn data were processed to produce calibrated event
lists. Unwanted hot, dead or flickering
pixels were removed, likewise events due to electronic noise,
and event energies were corrected for charge-transfer losses. The latest available
calibration files\footnote{{\tt http://xmm.vilspa.esa.es/ccf/}} were used
in the processing.  Light curves were extracted from these event lists to
search for periods of background flaring and showed the background to
be stable throughout the duration of the observation.  The total
amount of ``good'' exposure time selected was  42688~s for MOS1, 42692~s
for MOS2 and 38106~s for the pn. (The pn exposure time includes a
correction for out-of-time events in full-frame mode; Str\"{u}der \et
2001).

Source data were extracted from circular regions of radius 60\arcs\
for the MOS data, and radius 35\arcs\ for the pn.  (The smaller
extraction region was used in the pn in order to avoid the CCD chip
boundary $\sim40$\arcs\ from the pn aimpoint.)    The total number of
counts for the pn is 52377, for MOS1 and MOS2 14501 and 14634,
respectively.  For the pn camera, both source and background regions
were taken from inside the area corresponding to the hole in the pn
circuit board.  An examination of background images and spectra
confirmed  that the instrumental fluorescence radiation, especially
above 7~keV from Cu and Ni, is not affecting the source or background
spectra.  There are no signs of pile-up in the source region. Events
corresponding to patterns 0--12 were extracted from the MOS
data. Patterns 0--4 (single and double pixel events only) were used
for the pn analysis.

\section{Spectral analysis}
\label{sect:spectral}

The source spectra were grouped such that each spectral bin contains
at least 20 counts and were fitted using the XSPEC 11.0.1 software
package (Arnaud 1996). The latest publicly available responses were
used {\footnote{{ m1\_medv9q19t5r5\_all\_15.rsp and
m2\_medv9q19t5r5\_all\_15.rsp for MOS1 and MOS2, respectively and
epn\_ff20\_sdY9\_medium.rmf for the pn}}}. The spectra from the three
detectors were fitted simultaneously but with the relative
normalisations free to vary.  The quoted errors on the derived
best-fitting model parameters correspond to a 90 per cent confidence
level for one interesting parameter (i.e. a $\Delta \chi^{2}=2.7$
criterion) unless otherwise stated.  Values of $ H_0 = 70 $~km
s$^{-1}$ Mpc$^{-1}$ and $ q_0 = 0.5 $ are assumed throughout, and fit
parameters (specifically edge and line energies) are quoted for  the
rest frame of the source.

Fig~\ref{fig:raw_spec} shows the raw count spectra from the EPIC
cameras. The broad-band spectrum is steep. A simple power-law fitted
across the entire spectrum gives a slope $\Gamma \approx 3.8$, but
this fit is dominated by the spectrum below 1~keV where the statistics
are best.  The most striking spectral feature is the sudden drop by
a factor of more than 2 at around 7~keV, visible in both pn and MOS
spectra  (note that this is a lower limit as the energy resolution of
about 180~eV will smear a sharp edge).  The width of the feature is
not significantly larger than the energy resolution.  The response
matrices do not introduce any strong features around the observed
energy of this feature, nor do the cosmic or instrumental
backgrounds. In the 7--10~keV range the source is detected with a
significance level of about 20 $\sigma$, therefore  the spectral
feature is intrinsic to the source.  However, the limited photon
statistics above 10 keV do not allow us to  constrain the expected
recovery of the flux at higher energies.

In the absence of any known systematic effect that could introduce
such a feature into the spectrum, the rest of this paper discusses
possible physical origins of the feature assuming it to be intrinsic
to the source spectrum.  The much higher statistical weight at lower
energies (due to the strong soft excess emission and peak in the EPIC
response at $\sim 1$~keV) meant that fits to the full-band spectra
were dominated by the soft emission. In order to concentrate on the
feature at 7 keV the data below 2 keV were ignored in the following
analysis.

\begin{figure}
\hspace{0.5 cm}
\psfig{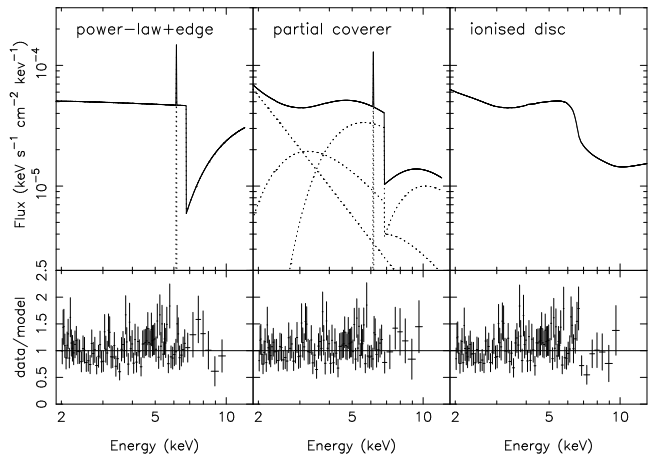}
      \caption{
Spectral fits to the 2--10~keV EPIC data.  The upper panels
demonstrate the models (in $F_{\nu}$ units) and the lower panels show
the data/model residuals.  The left panels show the simple power-law
and edge model, the middle panels show the partial covering model, and the
right panels show the $3\times$ solar Fe ionised reflection model.
\label{fig:fits}
}
\end{figure}

\subsection{Absorption by neutral iron?}

The most natural interpretation for a drop in the spectrum at just
above 7~keV is as an absorption edge from a large column of neutral
iron (7.112~keV, Henke \et 1993; 7.111~keV, Bearden 1964).  
The 2--10~keV spectrum (see Fig~\ref{fig:fits}) can be adequately parameterized ($\chi^{2} = 192 /
182 ~ dof$) using a power-law and a deep absorption edge ($\tau \rm =
1.8\pm0.3$).  While providing an acceptable fit to the data, this
model is unsatisfactory for a number of reasons. Firstly, the
2--10~keV power-law slope  is unusually flat ($\Gamma = 1.07\pm
0.09$).  Secondly, the steep spectrum below 1~keV suggests there is
little `cold' absorption above that expected from the Galactic column
($N_{\rm{H}}=5.8 \times 10^{20}$~cm$^{-2}$; Dickey \& Lockman 1990).
Indeed, the column of cold iron expected from the depth of the
putative edge is $N_{\rm{Fe}} \approx 6 \times 10^{19}$~cm$^{-2}$
which corresponds to  $N_{\rm{H}} \approx 2 \times 10^{24}$~cm$^{-2}$
for solar abundances.
At this column density the spectrum should be {\it dominated} by
absorption at low energies (L shell iron and K shell absorption from
lower $Z$ elements), whereas it actually steepens considerably in the
soft band.  Absorption by ionised material can produce a strong iron~K
edge without significant soft X-ray absorption, but the energy of the
putative edge suggests the absorber is not strongly ionised.  The edge
energy at $7.04\pm0.07$~keV is consistent with the neutral iron~K edge
energies given in the literature.  

The fluorescent yield for iron is close to 0.34 (Bambynek \et 1972).
However, there is no strong, narrow 6.4~keV emission line in the EPIC
spectrum; the 90 per cent upper limit on the photon flux in a narrow
6.4~keV line is $8 \times 10^{-7}$~photons cm$^{-2}$ s$^{-1}$
(corresponding to an equivalent width of $\sim 90$~eV).  The estimated
photon flux absorbed by the putative edge is $1.2 \times
10^{-5}$~photons cm$^{-2}$ s$^{-1}$.
For a spherically symmetric distribution of absorbing material one
expects a ratio of fluorescence line to Fe-K absorbed flux
approximately equal to the fluorescent yield, modified by the
absorption of the 6.4~keV line (which is about a factor of 2.5 here).
The measured ratio (of line/edge flux) is $\ls 0.07$, a factor $\gs 2$
smaller than expected for a spherical distribution.  This suggests
that the column density along the line-of-sight is higher than the
average surrounding the X-ray source. The solid angle subtended by the
putative absorber, as seen from the X-ray source, is likely to be $\ls
2\pi$~sr (but see also section~\ref{sect:disco}).

The following subsections discuss two alternative spectral models,
specifically, partial absorption by neutral material and reflection
from ionised material.

\subsection{Partial covering model}
\label{sect:partial_coverer}

One possible explanation for the 2--10~keV spectral form of \1 is
partial covering (i.e., a patchy absorber; Holt \et 1980): the
observed spectrum is modified by cold absorption at high energies,
while at lower energies a relatively unabsorbed component leaks
through. A partial covering model does provide a good fit to the
2--10~keV spectrum ($\chi^{2} = 177 / 178 ~ dof$), using three
absorbed power-laws to explain the flat spectrum while the deep iron~K
edge in the most heavily absorbed component explains the 7~keV feature
(see Fig~\ref{fig:fits}). A narrow 6.4~keV iron line is included in
the model although, as noted above, it is only poorly constrained by
the data.  The best fitting model parameters are as follows: intrinsic
power-law slope $\Gamma = 3.5_{-0.4}^{+1.2}$, line-of-sight column
densities (and covering fractions) of $N_{\rm{H}} \approx 0\ (0.04),
1.3\ (0.12)\ \rm and\ 4.6\ (0.84)$ (in units of $10^{22}~ \rm
cm^{-2}$) and an iron abundance of about 35.
Partial ionisation of metals could help, but the sharp spectral edge
suggests that iron is not strongly ionised and iron L-shell absorption
is unavoidable.  The edge energy in the partial covering model is
(7.10 $\pm$ 0.04) keV consistent with the energy of the neutral iron K
edge. Fitting the 0.3--10 keV
spectral energy distribution by adding a black body component, does not affect
the spectral parameters derived above.
\subsection{Ionised reflection model}
\label{sect:ion_disc}
A strong Fe edge may be found in the reflected emission from an
irradiated accretion disc which is not too highly ionised (Ross,
Fabian \& Young 1999). To investigate if such models can describe
these data, the constant-density reflection models of Ross \& Fabian
(1993) were fitted to the 2--10~keV spectrum. Since the reflected
emission may originate close to the central black hole, relativistic
blurring was applied to the model during fitting (using the {\tt LAOR}
code; Laor 1991). Reflection from ionised material was strongly
preferred over neutral reflection ($\Delta \chi^2 \approx -40$), an
iron abundance approximately 3 times solar was needed ($\Delta \chi^2
= -10$ from a solar abundance model), and a Kerr space-time geometry
was preferred over a Schwarzschild one ($r_{\mathrm{in}}$ = 3--4,
$r_{\mathrm{out}}$= 8.3 gravitational radii). However, a model with a
reflection fraction of unity can only give an adequate fit to the data
($\chi^2=258/180~ dof$) between 2 and 10~keV, and it cannot fully
account for the sharp spectral drop at $\sim$7~keV. Allowing the
reflection fraction to increase to 9 
resulted in a steeper continuum ($\Gamma=1.68$ from 1.54) and an
improved fit ($\chi^2 = 201/179 ~ dof$; see Fig~\ref{fig:fits}). 
The high value of $R$ can be decreased by increasing
the iron abundance of the reflector. The
ionisation parameter is $\xi \approx 750$~erg~cm~s$^{-1}$. At this
level of ionisation the reflection-dominated model predicts a strong He-like Fe~K$\alpha$
line with an equivalent width of 1.8~keV. This model clearly is not as
good at describing the data as the partial covering model. It is
extremely difficult for ionised reflection to account for the depth of
the drop at $\sim$7~keV without invoking a very extreme Fe abundance
and/or reflection fraction.

\subsection{Other models}

The blue wing of a very strong  relativistically broadened and
highly redshifted iron K$\alpha$ line can produce a drop at 7 keV. A
power-law plus {\tt LAOR} line (rest energy 6.4~keV) provides a good
fit ($\chi^{2} = 182 / 180 ~ dof$) with an inner disc radius
$r_{\rm{in}} < 2 r_{\rm{g}}$ and an underlying power-law slope of
$\Gamma=2.08_{-0.13}^{+0.25}$. However, the very high equivalent width
($EW=5$~keV) required to explain the size of the 7 keV drop is
difficult to justify. Alternatively, a reflection spectrum from cold
material will contain a neutral iron edge. Fitting with a {\tt PEXRAV}
model gives a good fit ($\chi^{2}= 192 / 182 ~ dof$) with a reflection
fraction $\gs200$ (essentially just reflection) and an iron
overabundance $>3$. The dominance of the reflected compared to the
primary emission, and the weakness of the associated iron K$\alpha$
line appear fatal for this interpretation.

\section{Temporal Analysis}
\label{sect:timing}

\begin{figure}
\psfig{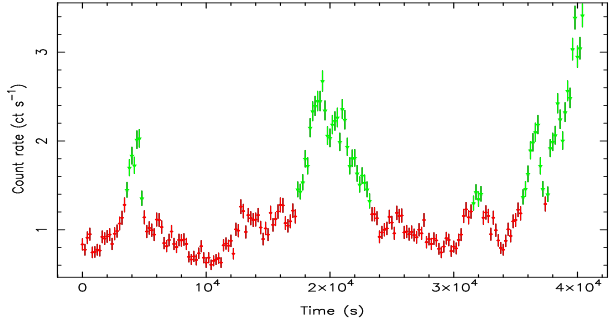}
      \caption{
EPIC pn light curve in the 0.1--10~keV energy range with a bin size of
200 s, demonstrating the strong and rapid variability. 
The
``high flux'' and ``low flux'' intervals  (see
section~\ref{sect:timing}) are marked differently.
\label{fig:light_curve}
}
\end{figure}

Fig~\ref{fig:light_curve} shows the 0.1--10~keV EPIC pn light curve in
200~s bins (The MOS light curves are essentially identical to the pn
light curve). The source again showed strong (factor of $\sim$4 change
during the observation) and rapid variability, as
previously seen by \asca\ (Leighly 1999a). The relatively low
apparent luminosity of the source during this observation, $L_{0.2-10}
\approx 8 \times 10^{42}$~erg s$^{-1}$, means that the rapid
variability translates to a fairly modest rate of change of
luminosity. For example, the rapid rise toward the end of the
observation corresponds to a luminosity increase of $\Delta L/\Delta t
\approx 2.4 \times 10^{39}$~erg s$^{-2}$. The corresponding radiative
efficiency, using the argument of Fabian (1979), is only $\eta \gs
0.2$ per cent.

\begin{figure}
\psfig{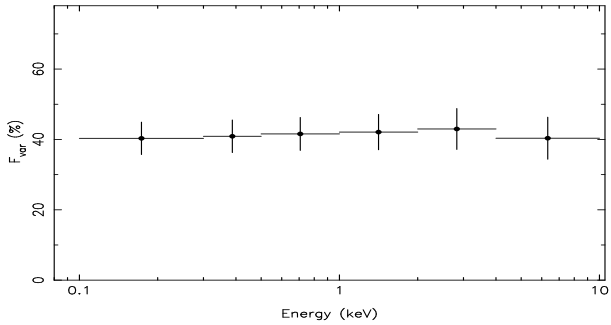}
 \caption{RMS variability versus energy for the EPIC pn light curve.
\label{fig:rms_spec}
}
\end{figure}

Light curves in various energy bands were analysed to search for
spectral variability. The variations in each band appear very similar.
The fractional variability amplitude $F_{\rm{var}}$ (see Edelson \et
2001) was measured in different energy bands (Fig~\ref{fig:rms_spec})
using light curves binned to 1000~s and is consistent with a constant
(fractional)  variability amplitude.  A cross-correlation analysis
using the discrete correlation function (DCF; Edelson \& Krolik 1988)
showed the light curves to be correlated at zero-lag, with no evidence
for any time lags.

\begin{figure}
\psfig{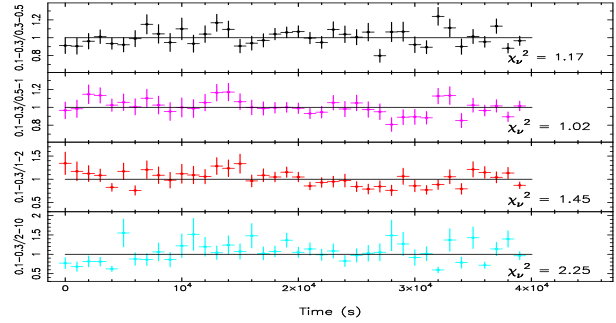}
 \caption{
Hardness ratios using 1000~s time bins. 
The first two hardness ratios (at the top) are acceptable fits to a constant.
The third is unacceptable at a $\sim$ 95 per cent level, the last one is
unacceptable at $>$ 99.9 per cent.  
\label{fig:hr}
}
\end{figure}

\begin{figure}
\psfig{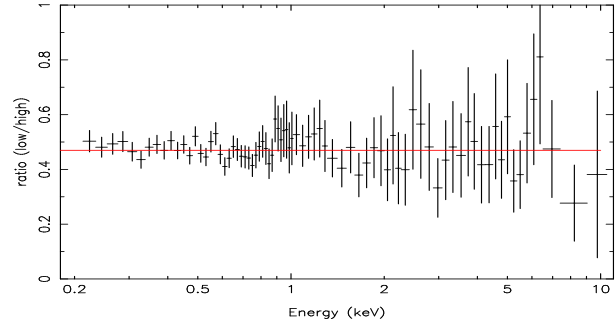}
 \caption{
Ratio of ``low flux'' and ``high flux'' spectra. \label{fig:ratio}
}
\end{figure}

Hardness ratios were also examined to search for spectral variability
(see Fig~\ref{fig:hr}). These do show spectral variability 
(the lower two panels)
but are uncorrelated with flux.
As a final check of (flux-correlated) spectral variability separate pn
spectra were extracted from time intervals when the source 
flux was below the mean (``low flux'' spectrum) and above the mean
(``high flux'' spectrum).
The ratio of the two spectra
(Fig~\ref{fig:ratio}) was consistent with a constant, re-enforcing
the claim of no flux-dependent spectral variability.  Unfortunately,
the limited number of counts above 7~keV ($\approx 140$ source
counts in the pn) mitigates against a detailed analysis of the
variability of the 7~keV spectral feature.

\section{Discussion}
\label{sect:disco}

The \xmm\ spectrum of \1 has revealed a sharp drop, by a factor $\gs
2$ in the spectrum at an energy just above 7~keV. This is to our
knowledge the first  detection of such a feature in an AGN. The
limited statistics and complex continuum mean that the identification
of this feature remains ambiguous. We have considered models involving
either partial covering of the X-ray source by cold material, or
emission from an ionised accretion disc, but both these explanations
have drawbacks.

\subsection{Comparison of the two models}

The energy of the feature is strongly suggestive of absorption by
neutral iron. However, a complex partial covering model is required in
order that the soft X-ray spectrum be relatively unaffected by the
absorber.  A further implication of the partial covering model is that
the underlying power-law is unusually steep ($\Gamma \approx 3.5$).
A similar feature, namely a sharp drop near 7.1~keV with relatively
little fluorescence emission, was observed in the peculiar X-ray
binary Cir X-1 (Brandt \et 1996).  In this case the spectral feature 
is almost
certainly due to partial covering.

An important consequence of the partial covering model is that the
unabsorbed X-ray luminosity of  \1 is an order of magnitude higher
than the observed (absorbed) luminosity.  We have calculated the $\rm
\alpha_{ox}$ values for the absorbed and unabsorbed fluxes at 2~keV
(the flux density at 2500~\AA\ is taken from Leighly 2000).  The
values are $\rm \alpha_{ox}\ =\ -2.0$ (absorbed X-ray flux) and $\rm
\alpha_{ox}\ =\ -1.6$ (unabsorbed). This is suggestive of absorption
(see Brandt, Laor \& Wills 2000; Brandt \et 2001, Figure 3) as almost
all the objects with $\rm \alpha_{ox}\ < \ -1.8$ suffer from heavy UV
and X-ray absorption. However, there is no evidence for intrinsic
absorption in the \hst-STIS spectrum of \1 (Leighly 2000).


The ionised disc model including relativistic blurring seems to
provide another possibility for the sharp spectral feature. The blue
horn of the relativistically broadened K$\alpha$ line emission,
combined with the iron edge can produce a drop around 7 keV.  However,
there still remain some residuals above 6.8 keV in the observers frame
and it appears somewhat surprising that the energy of the feature is
almost exactly that of the neutral iron edge.


\subsection{A physical scenario for the partial covering model}

The simplest configuration for the absorber would involve the absorber
completely covering our line of sight, the unabsorbed flux being a
scattered component from some other direction.  This scattered
component would not then vary rapidly and so would cause spectral
variability when combined with the (variable) absorbed flux. If on the
other hand the absorber does partially cover the source then it
presumably is at most only a few 100 light-seconds in size and its
density exceeds $\sim10^{12}$~cm$^{-3}$.
Photoionization considerations require that the cloud thickness be
less than $\sim 10^{12} r_{15}^{2}$~cm, where the distance of a cloud
from the nucleus is $r_{15}10^{15}$~cm.  The source position, size and
spectrum must remain fixed while the flux is rapidly varying in order
to avoid strong spectral variability.

These problems are alleviated if the absorbers are close to, or even
within, the emission region. The
absorbing clouds must then be very dense to remain cool and therefore
held by magnetic fields. The possible presence of such clouds in AGN in
general has been discussed by Rees (1987); Celotti, Fabian \& Rees (1992); 
Kuncic, Blackman \& Rees (1996); Kuncic, Celotti \& Rees (1997) 
and Malzac (2001). Brandt \& Gallagher (2000)  first suggested that such 
clouds might be particularly relevant for NLS1s.
Cool gas trapped by the magnetic field is compressed
to extreme densities by the high radiation pressure. Clouds and
filaments of such cold gas may accompany the active regions above an
accretion disc.

This possibility has been investigated using a
relativistically-blurred model involving emission and partial
covering. In order that the 7.1~keV feature remains sharp in the
model, the inclination must be $\ls 15\deg$. If the absorbing material
lies outside the emission region (say
in a toroidal structure at $10^{15}$~cm) then only the line-of-sight 
material
affects the spectrum and this constraint is relaxed.

It is interesting to speculate whether the absorbing material is
connected with the radiation-driven outflow suggested by Leighly
(2000) in \1, and whether it affects the observed rapid X-ray
variability. Clearly it is crucial to search for similar features in
other objects. The \xmm\ spectrum of PG~1211+143 (Reeves \et in prep.)
appears to show a similar drop at $\sim 7.3$~keV.  If such features
are common in other objects,  this will further constrain the range of
suitable models.
\section*{Acknowledgements}
Based on observations obtained with \xmm, an ESA science mission with
instruments and contributions directly funded by ESA Member States and
the USA (NASA). WNB acknowledges the financial support of NASA grant
NAG5-9939.
We thank Gareth Griffiths and Michael Freyberg 
for advice about the EPIC calibration and the anonymous referee for
constructive comments.

\bsp
\label{lastpage}
\end{document}